\documentclass[twocolumn, 10pt]{revtex4-1}

\usepackage{amsmath}
\usepackage{amssymb}
\usepackage{graphicx}
\usepackage{epsfig}
\usepackage{epstopdf}
\usepackage{color}

\begin{document}

\title{Spatial patterns of dissipative polariton solitons in semiconductor microcavities}

\author{J.~K.~Chana$^1$, M.~Sich$^1$, F.~Fras$^{1, 2}$, A.~V.~Gorbach$^3$, D.~V.~Skryabin$^3$, E.~Cancellieri$^1$, E.~A.~Cerda-M\'endez$^4$, K.~Biermann$^4$, R.~Hey$^4$, P.~V.~Santos$^4$, M.~S.~Skolnick$^1$ and D.~N.~Krizhanovskii$^1$}
\affiliation{$^1$Department of Physics and Astronomy, The University of Sheffield, Sheffield, S3~7RH, United Kingdom\\
        $^2$Institut NEEL CNRS, Grenoble, 38042, France\\
        $^3$Department of Physics, University of Bath, Bath, BA2~7AY, United Kingdom\\
        $^4$Paul-Drude-Institut f\"ur Festk\"orperelektronik, Berlin, 10117, Germany}

\begin{abstract}
Semiconductor microcavities operating in the polaritonic regime are highly non-linear, high speed systems due to the unique half-light, half-matter nature of polaritons. Here, we report for the first time the observation of propagating multi-soliton polariton patterns consisting of multi-peak structures either along (x) or perpendicular to (y) the direction of propagation. Soliton arrays of up to 5 solitons are observed, with the number of solitons controlled by the size or power of the triggering laser pulse. The break-up along the x direction occurs due to interplay of bistability, negative effective mass and polariton-polariton scattering, while in the y direction the break-up results from nonlinear phase-dependent interactions of propagating fronts. We show the experimental results are in good agreement with numerical modelling. Our observations are a step towards ultrafast all-optical signal processing using sequences of solitons as bits of information.
\end{abstract}

\maketitle


Solitons are perhaps one of the most exciting phenomena in nonlinear physics. They occur when dispersive spreading is compensated through non-linear interactions. Soliton formation has been demonstrated in numerous systems such as nonlinear crystals and optical fibers \cite{agr, amp} as well as atomic Bose-Einstein condensates \cite{becsol1}. Solitons provide an underlying mechanism for  processes in nature such as propagation of signals in neurons \cite{NerveSol}, cloud formation \cite{Baines:Topo} and large amplitude waves \cite{Russell, Apel:InternalSol}. Since solitons are robust against perturbations, they are also used for applications in long-haul communications \cite{Georges:640Gbit} as well as in all-optical on-chip signal processing schemes~\cite{Foster}.

In many aspects, solitons behave like artificial particles. Localised initial perturbations can split into two or more solitary waves~\cite{Slekys} resulting in the formation of multi-soliton complexes (molecules), in which solitons can dynamically move towards equilibrium spacing~\cite{Schapers}, or exhibit complex oscillatory behaviour~\cite{Scroggie}. Trains of bright conservative solitons have been reported in cold atom condensates and were attributed to repulsive interactions between solitons~\cite{becsol2}.

The first systems in nonlinear optics used to demonstrate the formation of stable nonlinear patterns and solitons were nonlinear resonators~\cite{Rosanov:Book}. In microcavities in the weak coupling regime the interplay between conditions leading to single soliton or pattern formation has been extensively studied \cite{Barland:CavitySol, Taranenko, Spinelli, LiquidCrystal}. Recently polaritons, hybrid light-matter particles forming in the strong coupling regime in semiconductor microcavities~\cite{book, book2}, have attracted significant interest. Non-linear hydrodynamic phenomena such as superfluidity~\cite{AmoNP2009} and both integer \cite{Nardin} and half vortices~\cite{Lagoudakis:HalfVort} have been observed in this system.  Importantly for applications in all-optical signal processing, polariton systems possess giant Kerr-like optical nonlinearity and can be manipulated on a picosecond timescale enabling the construction of miniature polaritonic circuits and logic gates~\cite{Cancellieri}.

Microcavity polaritons are an open system far from equilibrium. Recently bright polariton solitons have been observed~\cite{sich}. These bright spatial solitons exist when an external pump fully compensates photonic losses and the decay of the excitonic coherence and are therefore termed dissipative. Dark soliton trains arising from the break-up of polariton condensates in 1D conservative microcavity system (no pump) were recently predicted theoretically \cite{Pinsker:1}, whereas dissipative polariton soliton patterns in microcavities remain unexplored.

An understanding of soliton-soliton interactions is important not only for applications utilising solitons as information bits, but also to give an insight into the nonlinear properties of the polariton system.

In this article we study the fundamental mechanisms of polariton soliton formation and, for the first time, we demonstrate the formation of dissipative soliton patterns. The interplay between bistability of the external pump field, polariton-polariton scattering and polariton negative effective mass along the propagation direction enables the formation of x-soliton arrays, i.e. bound solitons following one after the other in space and time. Up to 5 stable bound solitons were observed, which is supported by numerical simulations. We also observed stable y-soliton arrays formed along the direction perpendicular to the propagation direction, which arise due to changes of the front velocity across the propagating beam profile.

\section*{Results}

In our work we used the same sample as described in Ref.~\cite{sich}, a GaAs-based $\lambda$-cavity with 6 GaAs quantum wells. The basics of soliton excitation are explained in Fig.~\ref{Fig:Basic}. In contrast to our previous work, the writing pulse (WP) which triggers soliton formation is chosen to be elongated ($\approx 5~\mu$m~$ \times~30~\mu$m) along (Fig.~\ref{Fig:Basic}(a)) or perpendicular to (Fig.~\ref{Fig:Array}(a)) the direction of soliton propagation. We name the resulting patterns in the two cases -- \emph{x-soliton arrays} and \emph{y-soliton arrays} respectively for ease of discussion. These are special cases of possible soliton patterns. X-soliton arrays are bound solitons moving one after another, while in the case of y-soliton arrays we observe multi-peak patterns along the transverse direction moving as a single front. All experiments reported here were performed in a single polarisation configuration with the pump and WP co-circularly polarised to minimise polarisation crosstalk~\cite{Sich:SolPol}.


\section*{X-soliton arrays}

\begin{figure}
\centering \includegraphics[width=8.5cm]{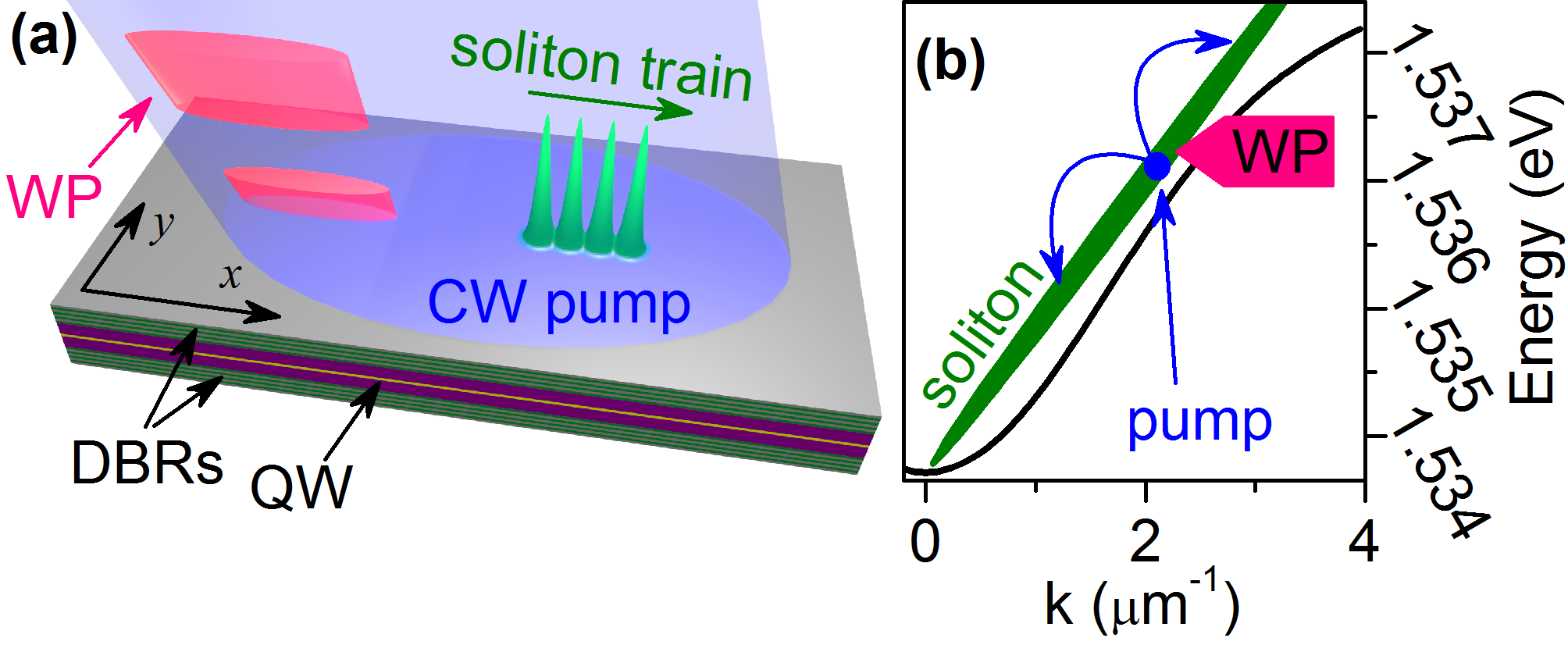}
\caption{\textbf{Basics of soliton excitation}. (a)~A semiconductor microcavity is a planar Fabry~-~Perot resonator with two distributed Bragg reflectors (DBRs) on resonance with excitons in quantum wells (QW). In the strong coupling regime, coupling between excitons and optical modes (photons) results in quasi-particles termed exciton-polaritons. In order to excite a soliton a large area of the sample is quasi-resonantly illuminated with a continuous wave (CW) pump. This maintains the system in the low density state within the bistability range and provides gain to sustain dissipative solitons. Then a resonant writing pulse (WP) (red) with a duration of a few picoseconds and focused into a small spot is used to trigger the soliton formation, locally switching the polariton system to the high density state, which then develops into a soliton thanks to negative effective mass and nonlinear polariton-polariton interactions. (b)~Polariton dispersion (solid black curve) of the lower polariton branch (LPB). The CW pump is slightly blueshifted relative to the LPB and the excitation angle is such that the injected polaritons have negative effective mass. Green -- schematic of the linear soliton dispersion resulting from the polariton-polariton scattering from the pump state and triggered by the WP.}
\label{Fig:Basic} \end{figure}

Bright polariton solitons are supported by a continuous wave (CW) pump resonant with the lower polariton branch at high momentum $k \approx 2~\mu$m$^{-1}$. The pump power has to be tuned into the bistability domain and the soliton can be considered as a local switch from the lower to the upper state on the bistability curve. Parametric polariton-polariton scattering from the pump populates soliton harmonics over a broad range of k-vectors. In order to excite x-soliton arrays, we apply the elongated WP just at the edge of the Gaussian CW pump spot (Fig.~\ref{Fig:Basic}(a)), i.e outside the region where the pump exhibits bistable behaviour. The WP is quasi-resonant with the lower polariton branch and its transverse momentum $k_{wp}$ and energy $\hbar\omega_{wp}$ are close to that of the pump: $k_{p} \approx 2.5~\mu$m$^{-1}$, $\hbar\omega_{p} \approx 1.54$~eV. Full widths at half-maxima (FWHM) of the WP are $30~\mu$m by $5~\mu$m along directions $x$ (soliton propagation direction) and $y$ (transverse direction) respectively, and the pulse duration is 5~ps.

\begin{figure}[t]
\centering \includegraphics[width=8.5cm]{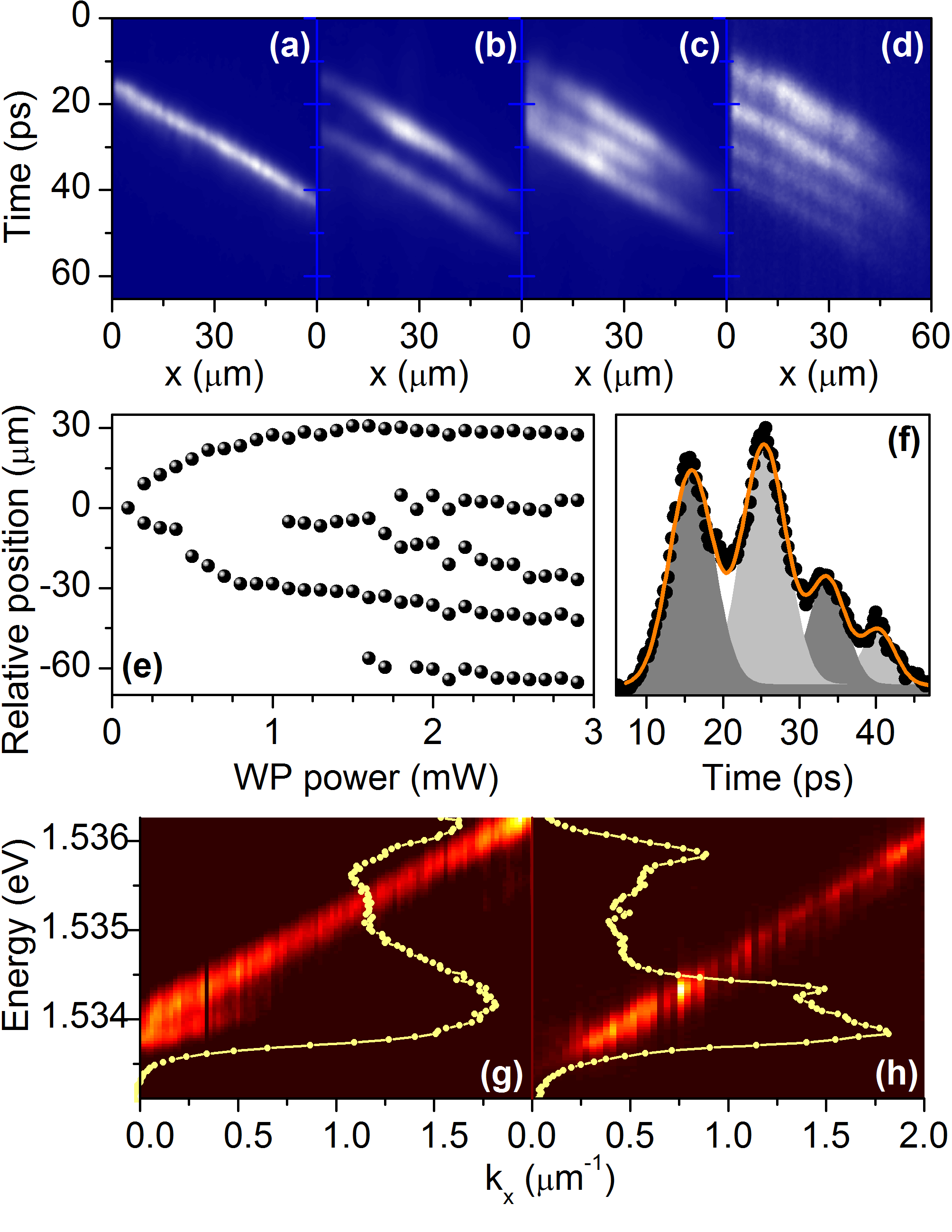}
\caption{\textbf{Experimental observation of x-soliton arrays}. (a)-(d) Streak camera images of a single soliton (a) and x-soliton arrays (b)-(d) for different WP powers; $t=0$ is the WP arrival time. For each time (vertical axis), the horizontal line shows the pseudo-colour intensity profile acquired along a narrow slit aligned with the $x$-axis of the system. Therefore, a soliton moving with a constant velocity appears as a tilted straight line on these images. (e) Relative positions and spacings of the centres of solitons as a function of the WP power. (f) Profile of a four-peak x-soliton array in time (as on (d)), where the average soliton duration is $\approx$~6~ps and the velocity is $\approx$~2.2~$\mu$m/ps (time is arbitrary and not related to the WP arrival time). (g),~(h)~Spectra of a single soliton and four-peak x-soliton array at $\approx~30$~ps; yellow lines show energy intensity distribution along the spectrum. }
\label{Fig:Train} \end{figure}

As we showed previously, polariton solitons are triggered at WP power densities above a certain threshold~\cite{sich}. At low WP powers only a small region in the centre of the Gaussian WP is expected to excite a single soliton whose size (healing length) is determined by the polariton-polariton interactions and the cavity parameters~\cite{sich}. At higher WP powers, the area across the injected polariton wavepacket where solitons may be excited becomes larger than the healing length, so multiple bound solitons are triggered.  Figs.~\ref{Fig:Train}(a)-(d) show spatio-temporal profiles of solitons for different WP powers. The number of created solitons increases with the WP power, as expected.  At first, at a WP power of 0.1~mW, only one soliton is excited (Figs.~\ref{Fig:Train}(a)), whereas at 0.2~mW two parallel soliton traces are observed (Figs.~\ref{Fig:Train}(b))). Notably, as the WP power increases, the spacing between solitons gradually increases and more soliton traces appear at higher WP power (Figs.~\ref{Fig:Train}(c, d, e)). The profile of the four-peak structure along the $x$-axis is shown in Fig.~\ref{Fig:Train}(f) with an average FWHM of a single peak of $\approx7$~-~$8~\mu$m, which is consistent with our previous observations. All four solitons travel at the same speed without crossing each other indicating formation of stable soliton patterns.

The important difference between solitons and dispersive wave packets is that the soliton dispersion is expected to be linear due to polariton-polariton interactions \cite{sich}. Experimental measurements of the energy vs momentum for the observed  single soliton and four-peak x-soliton arrays are shown in Figs.~\ref{Fig:Train}(g,~h). \footnote{Note, that we are only able to image k-vectors smaller than that of the pump because reflections of the pump and WP lasers located at higher k-vectors saturate our detectors.} In both case the spectra show linear dispersion of the soliton emission at all k-vectors down to zero in strong contrast to the dispersion of the lower polariton branch, which is parabolic at $k\approx0$ (Fig.~\ref{Fig:Basic}(b)). The yellow lines in Figs.~\ref{Fig:Train}(g,~h) show the intensity distribution of the soliton emission versus energy with two maxima at around 1.534~eV and 1.536~eV.  These peaks become narrower and more pronounced with an increase of the number of soliton humps, which has been predicted theoretically~\cite{Egorov:2D}. This narrowing can be understood in the following way -- the energy profiles are a Fourier transform of the temporal soliton pattern profile. For example, an infinite array of solitons can be represented by the interference between plane waves with different frequencies, and as a result its energy spectrum will be delta-function like, while in the case of a finite soliton array the peaks in the spectrum will have a finite linewidth. Overall, the k-space and real-space observations indicate the transition to the regime of parametric generation of a long sequence of solitons, theoretically studied in Refs.~\cite{Egorov:2D, Egorov:Param}.

We have carried out the same measurements and obtained very similar results for different values of $k_{wp}$ and energies, $E_{wp}$. The soliton spectrum is broad and is populated by parametric scattering from the pump state which can be triggered by a WP anywhere on the soliton dispersion, thus allowing soliton creation by a wide range of WP energies and k-vectors.

For numerical investigation of the system, we used Gross-Pitaevskii equations describing coupled TE and TM cavity modes interacting with the spin dependent polariton field \cite{sich} and looked for quasi-circularly polarised soliton solutions moving in the direction of the pump momentum (i.e. along the $x$-axis)~\cite{sich}. The model described in Ref.~\cite{sich} has proved to include all the important features necessary to numerically reproduce the experimental conditions of Refs.~\cite{sich, Sich:SolPol}, which are also used in the present work. The soliton existence range in terms of the pump intensity, $|E_{p}|^2$, is well approximated  by the bistability interval of the intracavity field \cite{sich}.

\begin{figure}[t]
\centering \includegraphics[width=8.5cm]{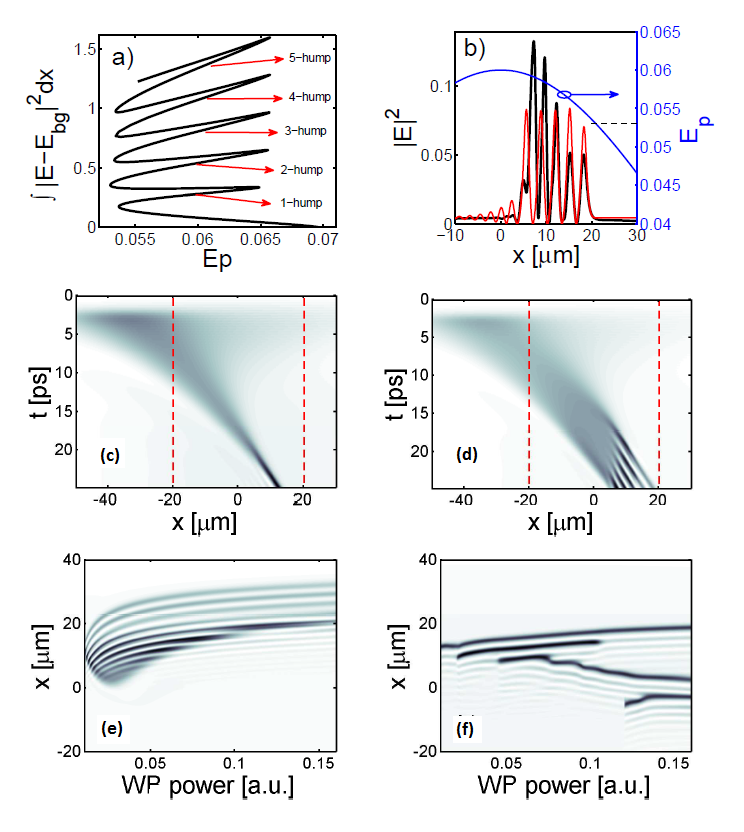}
\caption{\textbf{Numerical simulations of x-soliton arrays.} (a)~Snake-like bifurcation diagram for the multi-peak solitons in the case of infinite flat pump. $E(x)$ is the electric field amplitude, $E_{bg}$ is the amplitude of the homogeneous solution (in the tails of the soliton), and $E_{p}$ is the pump amplitude. (b) The red line is an example of the exact five-peak x-soliton array solution found for an infinite flat pump beam with amplitude $E_{p}=0.06$; the black line is the five-peak x-soliton array generated in the numerical modelling reproducing the experimental conditions of Gaussian pump profile, as in (e) and (f); the blue line shows the Gaussian profile of the pump field. (c),~(d)~Spatio-temporal dynamics resulting in the formation of the single-peak (c) and four-peak (d) x-soliton array. The pump field has a Gaussian profile with amplitude $E_{p}=0.06$, the corresponding boundaries of the bistability region (as in (a)) are indicated by vertical dashed lines. The WP amplitude is $E_{wp}=0.1$ (c), and $E_{wp}=0.12$ (d). (e),~(f) Profiles in pseudo-colour of multi-peak structures recorded in numerical simulations after 25~ps as a function of the WP power. The WP momentum exactly matches the pump momentum in (e) and is offset by $3^o$ in (f).}
\label{Fig:TrainsMod}
\end{figure}

We have computed the bifurcation diagram showing a sequence of stable multi-peak solitons (Fig.~\ref{Fig:TrainsMod}(a)). The diagram shows that for every given pump intensity (horizontal axis) within the bistability range, stable soliton solutions with different numbers of peaks are found depending on the WP power (vertical axis). An example of a five-peak x-soliton array is shown in Fig.~\ref{Fig:TrainsMod}(b). In our time-dependent modelling we assumed a Gaussian profile of the WP. With an increase of the WP intensity, there is an increase in the number of peaks in stable multi-peak soliton solutions, see Fig~\ref{Fig:TrainsMod}(c,d), in agreement with the experimental results on Fig.~\ref{Fig:Train}(a-e).

In practice, the profile of the pump beam is Gaussian and therefore conditions for the soliton formation are met only within a certain radius around the pump beam center, where the pump intensity is large enough to bring the cavity into the bistable regime.

In Fig.~\ref{Fig:TrainsMod}(b) we compare profiles found in the case of an infinite flat pump with those moving through the Gaussian pump. One can see that in both cases the peak widths and peak spacings are closely matched. The peak amplitudes, however, differ as the latter structure does not spend sufficient time inside the bistability range to fully stabilise itself.

Figs.~\ref{Fig:TrainsMod}(e,~f) show how the number and relative position of peaks varies with the WP power for the exactly matched and slightly offset momenta of the pump and WP beams respectively. Fig.~\ref{Fig:TrainsMod}(f) is in good agreement with the experimental observations on Fig.~\ref{Fig:Train}(e); there is an increase in the number of peaks with increasing WP power and the spacings change in an irregular way with increasing WP power, generally increasing at higher power. Minimising the difference between $k_{wp}$ and $k_p$ (preferably to zero) facilitates the generation of a regular pattern of peaks.

The solitons are formed by scattering polaritons from the pump state to populate the soliton spectrum. The characteristic time of polariton scattering is approximately given by $h/(gN)$, where $g$ is the interaction constant, $N$ is the density (then $gN$ is the blueshift). For a blueshift of 0.3~meV, about 13~ps are needed to fully populate the spectrum and form a soliton. The scattering can only occur once the trigger WP reaches the pump bistability area. For a Gaussian pump spot, solitons will only form after they have propagated a certain distance through the pump spot irrespective of how far away the WP is placed (Fig.~\ref{Fig:TrainsMod}(c,~d)).  For this reason, when the WP power is high enough to excite a x-soliton array, the solitons in the array tend to appear at approximately the same position relative to the pump bistability boundary, see Fig.~\ref{Fig:TrainsMod}(d).


\section*{Y-soliton arrays}

We note localisation along the $x$ direction occurs due to the interplay between the negative effective mass and repulsive interactions. However, we observe that solitons in the x-soliton train are also localised along the $y$ direction due to the interaction of propagating fronts combined with the phase-dependent parametric nonlinearity~\cite{Egorov:2D}. Therefore, we observe an array of two dimensional solitons, travelling one after another.

\begin{figure}[t]
\centering \includegraphics[width=8.5cm]{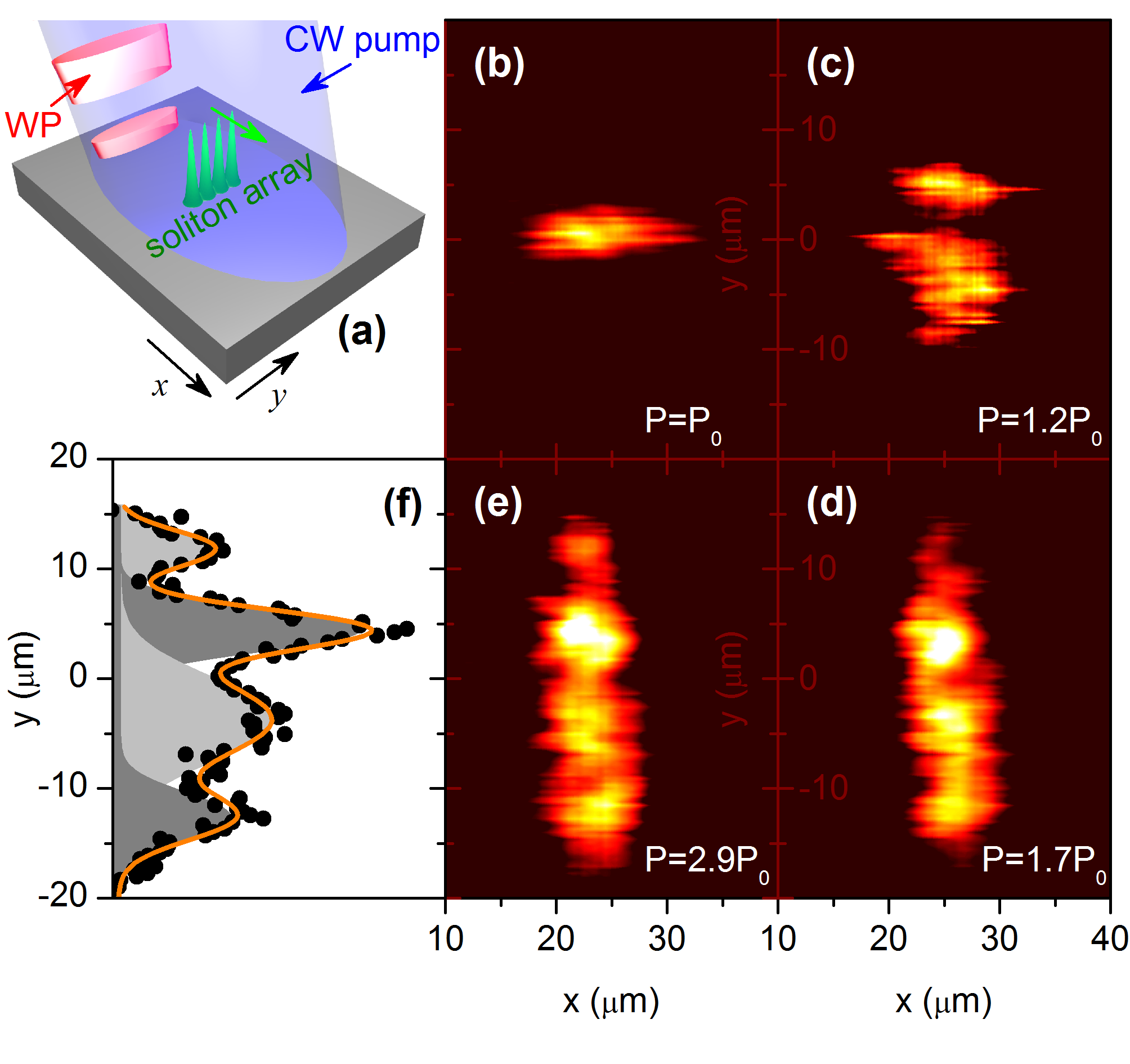}
\caption{\textbf{Experimental observation of y-soliton arrays.} (a)~Experimental setup with the WP elongated along the $y$ axis. (b)-(e) 2D images of one- to four-peak arrays obtained for different WP powers, $P_0 = 230$~$\mu$W. (f) Profile of a four-peak array along the $y$ axis.} \label{Fig:Array}
\end{figure}

In the second set of experimental measurements we investigated how localisation along the \emph{y} direction affects pattern formation along this direction. We elongate the WP along the $y$ direction, so it is  $\approx7~\mu$m along the $x$-axis and $\approx30~\mu$m along the $y$-axis. In this case we observe breaking of the initially smooth beam profile into a y-soliton array, see Fig.~\ref{Fig:Array}.

As with the x-soliton arrays the formation of the patterns only occurs in the soliton regime. The same experiment when the pump power is below the bistability regime never leads to formation of stable patterns, but to a dispersive wavepacket.

The number of peaks created depends, as for the x-soliton arrays, on the effective spot size of the WP. Figs.~\ref{Fig:Array}(b)-(e) show real-space images of the emerging structures for different WP powers. Peaks are spaced by $\approx8~\mu$m, are of the size (FWHM) of $\approx7~\mu$m (see Fig.~\ref{Fig:Array}(f)) and travel along the $x$ direction within the pump bistability area.

Similar to the x-soliton array, increasing the power leads to the creation of additional peaks on the 'sides' of the WP. It is important to note that the location of solitons triggered in such a way does not depend on the position on the sample: moving the sample by 10~$\mu$m relative to the laboratory frame does not influence the shape of the array. At the same time moving the WP along the $y$ axis relative to the pump and sample, which are fixed in the laboratory frame, leads to the respective shift of the soliton array trajectory.

\begin{figure}[t]
\centering \includegraphics[width=4.cm]{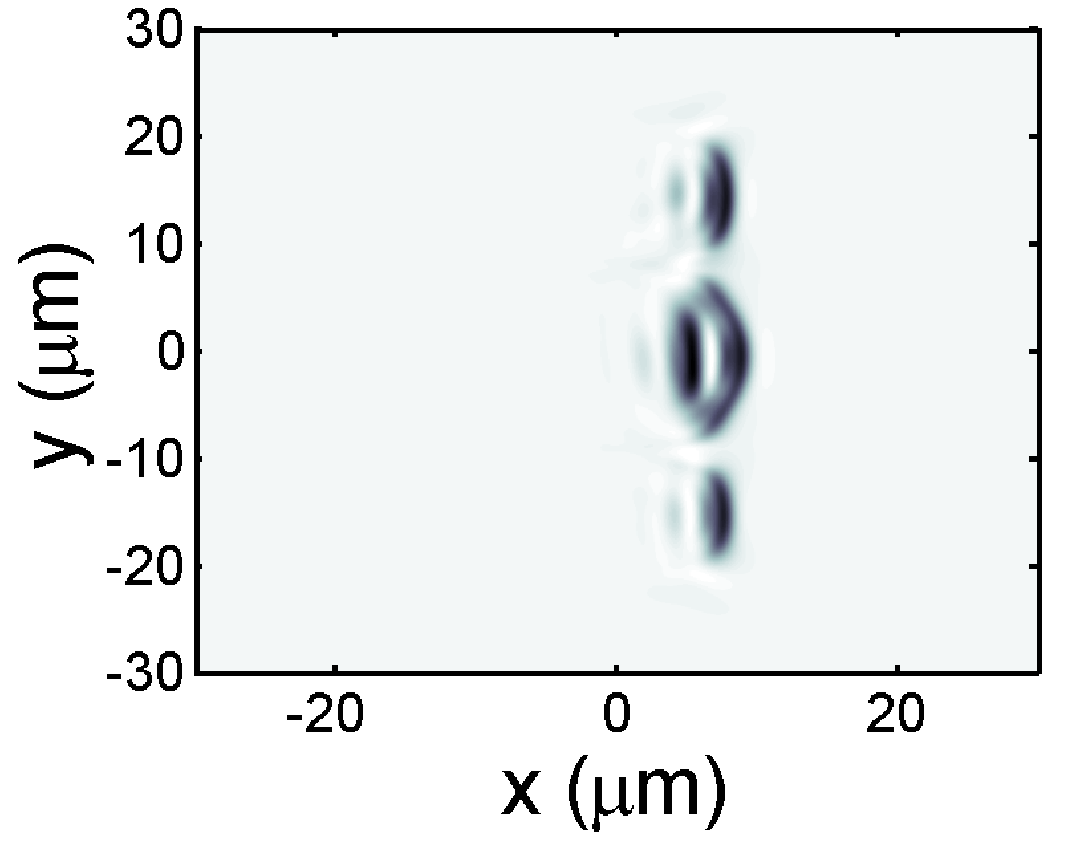}
\caption{\textbf{Numerical modelling of y-soliton arrays.} Numerically modelled transverse distribution of the intracavity optical intensity after 15~ps of propagation under the conditions when the WP is elongated along the $y$-coordinate (as in the experimental results shown in Fig.~\ref{Fig:Array}) and 3 degrees angle between the WP and pump momenta. Note that this figure is obtained for a single WP phase, while in the experiment we average over all possible WP phases. After averaging phases in the numerical modelling results are broadened, similar to those seen experimentally.}
\label{Fig:ArrayMod}
\end{figure}

In order to explain the above observations, we have first taken a single hump one dimensional soliton localised along the $x$ direction and infinitely extended along $y$ and performed its linear stability analysis. This demonstrated that it was stable with respect to any transverse instabilities, which could lead to filamentation along the $y$ direction. If, however, we limit the extent of the soliton stripe by imposing a Gaussian profile in the $y$ direction, then the change in the curvature of the soliton front leads to changes of the front velocity across the beam profile. This in turn leads to the breakup of the wavefront and the formation of y-soliton arrays, as in experiment. The latter process develops faster and is more pronounced if a small angle between the pump and WP beams is introduced (see Fig.~\ref{Fig:ArrayMod}).


\section*{Conclusions}

In the present study we have shown the existence and evolution of dissipative multi-soliton structures. X-soliton arrays and y-soliton arrays can be triggered by elongating the WP along the $x$ and $y$ directions respectively. Increasing the WP effective size (by changing its power) leads to the creation of larger structures with more peaks. X-soliton arrays can be excited across a broad range of initial conditions. This feature is typical for dissipative systems, where the soliton shape and energy are predetermined by the system parameters rather than the initial conditions.

We note that the physical mechanisms underlying break-up in the $x$ and $y$ directions are very different. The formation of y-soliton arrays results from inhomogeneity across a wavefront in the transverse direction. In contrast, localisation and break-up in the $x$ direction manifests a transition from a single soliton regime towards multi-peak parametric generation. An interesting extension of the study could be to use a WP which is large in both the $x$ and $y$ directions to observe the formation of more complex solitonic structures, although we have not yet been able to excite such structures due to lack of power density of the WP when it is spread over a large area.

Finally we note that from the applied point of view this study sets the framework for analysis of soliton-soliton interactions in digital processing devices where polariton solitons are used as data bits. Given the time separation between solitons these devices are expected to operate at speeds equivalent to a few hundred GHz clock rate.

\section*{Acknowledgements}

We acknowledge support from the Leverhulme Trust, EPSRC grant number EP/J007544/1, ERC Advanced Grant EXCIPOL 320570.

\section*{Corresponding authors}

D.~N.~Krizhanovskii (d.krizhanovskii@sheffield.ac.uk) and M.~Sich (m.sich@sheffield.ac.uk)

\end{document}